\documentclass[conference]{IEEEtran}
\usepackage{graphicx}
\usepackage{amsmath}
\usepackage{cite}
\usepackage{hyperref}
\usepackage{xcolor}
\usepackage{booktabs}

\title{Finetuning Strategies for Querying Sounds by Vocal Imitation}

\author{
\IEEEauthorblockN{Aditya Bhattacharjee, Christos Plachouras, Sungkyun Chang, Emmanouil Benetos}
\IEEEauthorblockA{
    School of Electronic Engineering and Computer Science, Queen Mary University of London, UK
}
}

\begin{document}
\bstctlcite{IEEEexample:BSTcontrol}

\maketitle

\begin{abstract}
    This technical report describes our winning submission to the AES AIMLA 2025 Challenge on querying sound effects by vocal imitation. We investigate two complementary fine-tuning strategies: contrastive learning with a frozen, pretrained CED encoder, and joint contrastive-triplet learning with semi-hard negatives using a MobileNetV3 encoder. This report has been updated for posterity to include details released after the challenge. 

\end{abstract}

\section{Introduction}

Query by Vocal Imitation (QbVI) involves retrieving audio samples from a reference collection based on a vocal imitation provided by the user. This problem is particularly challenging due to the variability in human vocalizations and the acoustic diversity of sound classes. Effective systems must learn representations that bridge the modality gap between vocal and non-vocal audio, while also being robust to the diverse ways humans imitate sounds.

Early approaches to QbVI focused on latent bottleneck features from autoencoders~\cite{zhang2016supervised} or semi-Siamese convolutional architectures trained with contrastive loss~\cite{zhang2017iminet}. These methods typically relied on hand-crafted or moderately learned features, combined with traditional classifiers or similarity measures such as dynamic time warping or cosine distance.

More recently, Greif et al.~\cite{greif2024improving} demonstrated the effectiveness of contrastive learning with neural audio embeddings pre-trained on large-scale datasets like AudioSet. Their system employed a dual-tower MobileNetV3 architecture, fine-tuned using SimCLR-style contrastive objectives and an extensive augmentation pipeline. This approach set a strong baseline for neural QbVI systems.

In this work, we focus specifically on neural network-based representations for QbVI and explore two complementary submission strategies for the AES-AIMLA 2025 Challenge. Our first submission involves contrastive fine-tuning of a frozen Consistent Ensemble Distillation (CED) encoder~\cite{dinkel2024ced}, which was originally trained for audio tagging via knowledge distillation. Our second submission builds upon the MobileNetV3 architecture and the setup of Greif et al., but integrates a triplet-based regularization objective. The motivation is to improve generalization by leveraging hard or semi-hard negatives sampled from a curated unpaired set of vocal imitations.\footnote{Submission \#2 is deployed on our website: \url{https://thatsoundslike.me}.}

\section{Datasets}

We use two primary datasets for training:

\begin{itemize}
    \item \textbf{VimSketch Dataset} \cite{pishdadian2019classifying}: This dataset provides supervised (reference, imitation) pairs with known correspondence. These pairs are used directly in our contrastive learning objective.
    \item \textbf{VocalSketch Dataset} \cite{cartwright2015vocalsketch}: This dataset contains vocal imitations that were excluded from the curated VimSketch set. Although they do not have matched reference samples, they are labeled by sound class. We use these as structured negatives in a triplet learning framework, sampling same-class negatives when available and falling back to random sampling otherwise.
\end{itemize}

In our first submission, only the VimSketch dataset is used. The second submission incorporates both datasets to enable a hybrid training regime that combines contrastive and triplet learning.

We use an online augmentation methodology on both query and reference data. This strategy is used in both our submissions to improve robustness to variation in vocal imitations and reference recordings. At each training step, a random subset of transformations is applied to each audio sample.

The augmentations include time-domain operations such as time shifting, gain adjustment, pitch shifting, and time stretching. In addition, we incorporate light additive Gaussian noise and frame-level corruptions to simulate temporal dropouts and distortions common in human vocalizations. Specifically, each sample is passed through up to one transformation randomly selected from a pool that includes:

\begin{itemize}
\item \textbf{Shift:} Random temporal offset of the waveform within $\pm$30\% of its length.
\item \textbf{Gain:} Random amplification or attenuation between $-10$ and $+10$ dB.
\item \textbf{Pitch Shift:} Semitone shift within a range of $\pm3$ semitones.
\item \textbf{Time Stretch:} Temporal scaling by a factor between $0.8$ and $1.2$.
\item \textbf{Additive Gaussian Noise:} Injects low-amplitude noise with amplitude sampled between $0.001$ and $0.015$.
\item \textbf{Frame-Level Corruption:} Applies stochastic frame-wise duplication, silencing, or removal to simulate human inconsistency.
\item \textbf{Frequency and Time Masking:} Randomly masks contiguous frequency bins or time frames in the spectrogram to increase robustness to partial input corruption.
\end{itemize}

This online augmentation is applied independently to both the query and reference branches of the model during training. It plays a crucial role in improving generalization across different vocalization styles, recording conditions, and intra-class variation.

For our validation dataset, we use the official \texttt{qvim-dev} set provided as part of the QVIM 2025 Challenge. Evaluation is performed using the Mean Reciprocal Rank (MRR) and Normalized Discounted Cumulative Gain (NDCG) metrics.

\section{Submission \#1}

\subsection{Architecture and Training Setup}

Our first submission is based on the Consistent Ensemble Distillation (CED) framework, originally introduced by Dinkel et al.~\cite{dinkel2024ced}. CED is a ViT-based audio encoder trained for AudioSet audio tagging using consistent ensemble distillation. It performed strongly on the HEAR 2021 benchmark, particularly for vocal imitation classification tasks, a motivation for its inclusion here. 

In this submission, we freeze the pretrained CED-base encoder (768-dimensional output) and train a lightweight MLP projection head to map embeddings to a 256-dimensional space suitable for retrieval. The encoder is shared across both the query and reference branches. Audio is resampled to 16 kHz and processed using the pretrained CED feature extractor. We extract the final encoder hidden representation, average-pool it over the sequence dimension to obtain a 768-dimensional embedding, and keep the CED encoder frozen during training.

The training objective is a supervised contrastive loss applied to known (reference, imitation) pairs from the VimSketch dataset. In-batch negatives are used to structure the embedding space. 

We apply online data augmentation symmetrically to both the query and reference audio inputs. Each training example receives a randomly selected transformation from a curated pool of time-domain effects; namely, the ones discussed in Section II. 

\subsection{Implementation}

The system uses the AdamW optimizer with a cosine learning rate schedule (starting at $1 \times 10^{-3}$), weight decay of $1 \times 10^{-5}$, and a batch size of 128. The model is trained for 50 epochs, with evaluation performed after every epoch. The temperature $\tau$ for the contrastive loss is fixed at 0.07. The output projection is a linear layer mapping the encoder's 768-dimensional output to a 256-dimensional embedding, which is then used to compute cosine similarities for retrieval.

\section{Submission \#2}

\subsection{Architecture and Training Setup}

Our model architecture builds on the QbVI framework proposed by Greif et al.~\cite{greif2024improving}, using a MobileNetV3 encoder pretrained on AudioSet~\cite{gemmeke2017audio}. In contrast to the dual-tower encoder architecture of~\cite{greif2024improving}, we use a single shared encoder for both reference and imitation audio, promoting tighter alignment in the learned embedding space and reducing parameter overhead. The encoder outputs are L2-normalized, enabling cosine similarity as a metric for both the loss computation and retrieval.

We adopt a hybrid learning strategy where supervised contrastive learning is regularized by a triplet loss. Contrastive loss is applied over positive pairs sampled from the VimSketch dataset, while the triplet loss uses negative examples drawn from the excluded subset of the VocalSketch dataset. These excluded examples consist of practice recordings and human-rejected imitations and do not participate in standard contrastive supervision. For triplet construction, negatives are sampled based on shared AudioSet ontology labels where available, falling back to random negatives otherwise. A semi-hard mining strategy is used, selecting negatives that are closer to the anchor than easy negatives but still harder than positives.

To adaptively balance the contribution of both losses during training, we implement a dynamic weighting strategy where the weight assigned to the triplet loss is scaled based on the number of active semi-hard triplets in the batch. If a few valid triplets are found, the triplet loss is down-weighted to avoid destabilizing training.

\subsection{Implementation}
The input features closely follow AudioSet’s preprocessing pipeline, where all audio is resampled to 32\,kHz and converted to 128-band log-Mel spectrograms using a 10-second duration, a window size of 800 samples, a hop size of 320, and an FFT size of 1024. Online data augmentation, as described earlier, is applied independently to both the query and reference inputs during training to increase robustness to vocal variation and noise. 960-dimensional embeddings are output from the penultimate layer of the encoder; the classifier head of the encoder is discarded. 

We train using the AdamW optimizer and a cosine learning rate schedule with one warmup epoch. The maximum learning rate is set to $2 \times 10^{-4}$ and the minimum to $5 \times 10^{-5}$. Training runs for 30 epochs with a batch size of 64. We fix the contrastive temperature to $\tau = 0.07$ and the triplet margin to 0.6. 

\section{Results}

The evaluation follows the official QVIM 2025 validation protocol, using the `qvim-dev` set provided by the organizers. The validation set includes both query-reference pairings and class metadata, enabling both pair-wise and class-wise evaluation.

The primary evaluation metric is the Mean Reciprocal Rank (MRR) computed over known query-reference pairs. Secondary metrics include Normalized Discounted Cumulative Gain (NDCG) and class-wise MRR, which assess ranking quality and class-level retrieval consistency, respectively.

Table~\ref{tab:results} summarizes the validation performance for both of our submissions, in comparison to the baselines. For posterity, we report that Submission \#2 is the winning submission in the AES AIMLA 2025 Challenge on querying sound effects by vocal imitation. Further details about the objective and subjective evaluation on test datasets can be found in the challenge website \footnote{\url{https://qvim-aes.github.io/}}. 

\begin{table}[htbp]
\centering
\caption{Performance Metrics}
\label{tab:results}
\begin{tabular}{lcc}
\toprule
\textbf{Model Name} & \textbf{MRR$_{ex}$} & \textbf{NDCG$_{cat}$} \\
\midrule
random       & 0.0444 & 0.337 \\
2DFT        & 0.1262 & 0.4793 \\
Greif et al.  & 0.2726 & 0.6463 \\ 
\midrule
Submission \#1     & 0.2876 & \textbf{0.6600} \\
Submission \#2     & \textbf{0.2932} & 0.6468 \\

\bottomrule
\end{tabular}
\end{table}

\section{Conclusion}

The goal of the Query by Vocal Imitation (QbVI) task is to retrieve audio samples from a reference database based on a vocal imitation provided by the user. As part of our participation in the AES-AIMLA 2025 Challenge, we explored two complementary approaches aimed at improving neural-network-based retrieval systems for QbVI. The submission differ in encoder architecture, training strategy, and supervision design.

In Submission~\#1, we leverage a frozen CED-base encoder pretrained via knowledge distillation on AudioSet. A supervised contrastive loss is applied over known reference-imitation pairs, using in-batch negatives to structure the embedding space. This lightweight approach achieves strong retrieval performance with minimal fine-tuning. In Submission~\#2, we extend the training pipeline by fine-tuning a shared MobileNetV3 encoder using a hybrid loss combining contrastive and triplet objectives. Semi-hard negative mining is performed using excluded vocal imitations, which are sampled based on class labels. A dynamic weighting strategy balances the two losses based on the availability of active triplets.

Together, these submissions highlight the value of both transfer learning and curriculum design in QbVI systems. Whether through careful pretraining or structured negative sampling, both strategies yield significant gains in retrieval performance over baseline systems.

\bibliographystyle{IEEEtran}
\bibliography{references}

\end{document}